\documentstyle[epsf]{Rspublic}  
\input{psfig}
\begin{document}   
\date{ }   
\author[Jo\~ao Magueijo, Kim Baskerville] {Jo\~ao Magueijo$^1$ 
and Kim Baskerville$^2$}  
\title[Big Bang riddles and their revelations]  
{Big Bang riddles and their revelations}  
\affiliation{$^1$ Theoretical Physics, The Blackett Laboratory,  
Imperial College, Prince Consort Road, London, SW7 2BZ, U.K.\\ 
$^2$ Department of Mathematical Sciences, South Road, Durham, 
DH1 3LE, UK.}  
   
\maketitle

\begin{abstract}{Cosmology; Inflation; String theory; Varying constants}  
We describe how cosmology has converged towards a beautiful model  
of the Universe: the Big Bang Universe. We praise this model, but show  
there is a dark side to it. This dark side is usually called ``the  
cosmological problems'': a set of coincidences and fine tuning features  
required for the Big Bang Universe to be possible. After reviewing these  
``riddles'' we show how they have acted as windows into the very early   
Universe, revealing new physics and new cosmology just as the Universe   
came into being. We describe inflation, pre Big Bang, and   
varying speed of light theories. At the  
end of the millennium, these proposals are seen respectively 
as a paradigm, a  
tentative idea, and outright speculation.  
\end{abstract}   
  
\section{The Big Bang riddles}  
  
The Big Bang Universe is a success story. It makes use   
of the general theory of relativity to set up the most minimalistic   
model for our Universe. According to this model the embryo Universe was   
concentrated in a single point, which exploded in a Big Bang event some 15    
billion years ago. The Big Bang Universe is homogeneous in space,    
and expands as time progresses: a dynamical prediction of relativity. An    
elegant explanation for an ever growing array of observations ensues.    
   
A closer examination of this    
model, however, reveals a number of unnatural features.   
The Big Bang Universe is fragmented into many small regions, 
which are so far apart light, or indeed any interaction,  
has not had time to travel between 
them. These ``horizons'' 
are therefore unaware of each other, yet mysteriously    
share the same properties, such as age and temperature. It   
almost looks as if telepathic communication has taken place  
between disconnected regions. Another puzzle  
is the observed near flatness of the Universe. Flatness is   
central to successful Big Bang models, but  
is unfortunately not stable.    
Big Bang models may be open  
(hyperbolic), flat, or closed (spherical). Closed   
Big Bang models expand to a maximum size, and then recollapse, dying   
in a Big Crunch. Open models expand too fast,   
leaving the Universe empty soon after the Big Bang.   
The problem is that even   
slight deviations from flatness grow very quickly,   
leading inevitably to either a catastrophic Big Crunch or emptiness.   
The fact that neither has occurred means that we are  
successfully walking on a tightrope. Short of invoking  
divine intervention how can we possibly have managed this  
for so long?  
  
Thankfully a number of natural explanations   
have been put forward. In all of these, the riddles plaguing the Big Bang   
act as windows into new physics. Inflationary models of the Universe,  
which have become a paradigm in modern cosmology, were undoubtedly born   
out of these puzzles. Inflation is perhaps the simplest addition to the   
Big Bang which leaves behind a Universe without mystery. Another   
explanation is the so-called Pre Big Bang model. This is inspired   
by string theory, and explores the possibility of the Universe existing   
before the Big Bang. In the progenitor Universe lies the secret of the  
riddles. The most radical explanation is a recent proposal, involving   
a revision of the special theory of relativity. According to this proposal  
light might have travelled much faster in the Early Universe.    
The varying speed of light cosmology explains the puzzles solved   
by inflation and Pre Big Bang models, and maybe some additional 
riddles, too. 
 
In this paper we review the Big Bang model (Section~\ref{bb}), 
and its riddles (Section~\ref{bbp}). We then describe solutions 
to these riddles: inflation (Section~\ref{infl}), Pre Big Bang 
(Section~\ref{pre-bb}), and varying speed of light (Section~\ref{vsl}). 
We conclude with an assessment of the state of the art.

\section{The bright side of Big Bang cosmology}  \label{bb} 
Cosmology, the study of the Universe, was for a long time the    
subject of religion. That it has become a branch of physics is    
a surprising achievement. Why should a system as apparently complex as    
the Universe ever be amenable to scientific scrutiny?   
At the start of this century, however, it became obvious that   
in a way the Universe is far simpler than say an   
ecosystem, or an animal. In many ways even a suspension bridge is far   
harder to describe than the dynamics of the Universe.    
   
The big leap occurred as a result of the discovery of the theory   
of relativity, in conjunction with improvements in astronomical observations.   
If we look at the sky we see an overwhelming plethora of detail:   
planets, stars, the Milky Way, the nearby galaxies. At first the    
task of predicting the behaviour of the Universe as a whole   
looks akin to predicting the world's weather, or the currents in the   
oceans.    
   
If we look harder we start to see that such structures are mere   
details. With better telescopes we can zoom out, to find that   
galaxies, clusters of galaxies, even 
the largest structures we can see, become    
``molecules'' of a rather boring soup. A very homogeneous soup,    
called the cosmological fluid. The subject of cosmology is the dynamical   
behaviour of this fluid, when left to evolve according to its own   
gravitation. The crucial feature is the fact that this fluid   
appears to be expanding: its ``molecules'' are moving away from each other.    
   
What set the Universe in motion?  
Can physics explain this phenomenon? That was one of the many historical   
roles played by the theory of relativity. The result is encoded in what came    
to be known as the Big Bang model of the Universe. Here we shall attempt   
to convey the essence of this model, in a Newtonian version of the   
theory which imports all the relevant relativistic aspects.    
  
\begin{figure}  
\centerline{\psfig{file=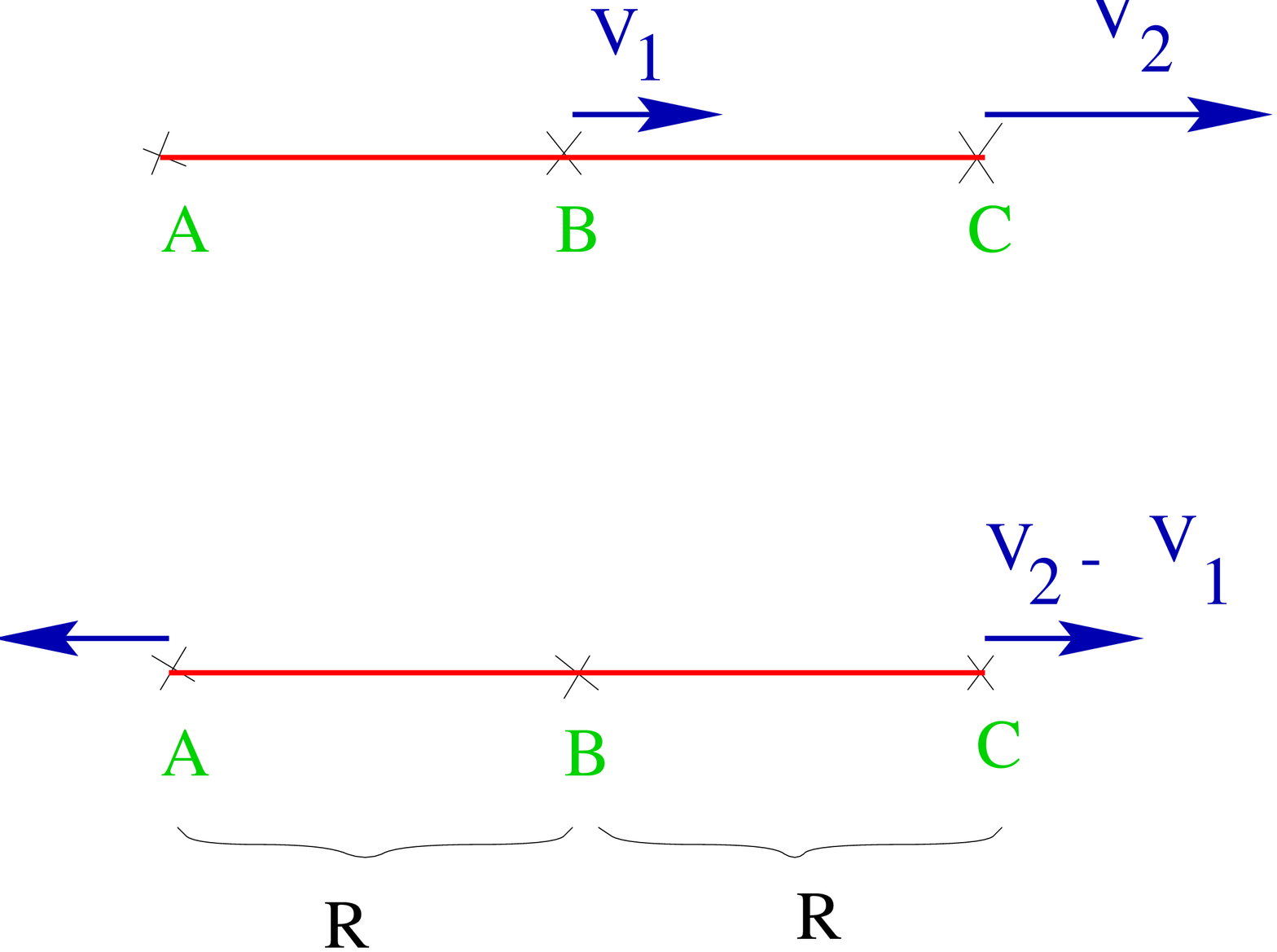,width=6 cm,angle=-90}}  
\caption{}  
\label{fig0}  
\end{figure}     
Let us start by assuming that the cosmological fluid is homogeneous   
and also that at every point all directions are equivalent, that   
is we have isotropy (note that homogeneity does not imply isotropy.)  
Isotropy requires that the only possible motion relative to any  
given point ${\cal O}$ be radial motion. Imagine a sphere around  
${\cal O}$, and consider the velocity vectors of points on this  
sphere. Now try to comb this sphere (that is, to add a tangential  
component to these velocities). There will always be a bald patch,   
no matter how careful one is. Such a bald patch   
provides a preferred direction, contradicting isotropy. Therefore  
isotropy is a hair raising experience, allowing only radial motion.   
Any observer can at most see outwards or inwards motion around itself.  
We shall assume for the rest of the argument that this motion is outwards.  
   
The speed at which this motion takes place may depend on    
the distance and on time, but the function must be the same for    
all central points ${\cal O}$ considered.    
This imposes constraints on the form of this function.   
Consider three collinear points $A$, $B$ and $C$, with $B$    
at distance $R$ from $A$ and $B$ (see Fig.\ref{fig0}).    
In the rest frame of $A$, let the velocities of $B$ and $C$ be    
${\bf v}_1$ and ${\bf v}_2$ (top case in Fig.~\ref{fig0}).  
If we now  consider the situation from the point of view   
of $B$, $C$ is at a distance $R$ and has velocity   
${\bf v}_2-{\bf v}_1$ (bottom case in Fig.~\ref{fig0}).   
But given homogeneity, what $B$ now   
sees at $C$ should be what $A$ sees at $B$. Hence   
${\bf v}_1={\bf v}_2-{\bf v}_1$, that is ${\bf v}_2= 2{\bf v}_1$.   
Going back to $A$'s perspective, we now find that points   
at twice the distance move at twice the speed. More generally   
${\bf v}=H {\bf d}$: the recession speed away from any point   
${\cal O}$ is proportional to the distance. $H$   
is called the Hubble ``constant''. It is not really a constant, but   
may only depend on time.   
  
This is a weird law. Any point ${\cal O}$  
sees the stuff of the Universe receding  
away from it, the faster the further away it is.  
Let us first simplify life, and  
ignore gravity, so that these speeds do not change in time.   
Then a cataclysm must have happened in the past. If an object at   
distance $d$ is moving at speed $v=Hd$, then rewinding the film by   
$\delta t= d/v=1/H$ will show that this object was ejected from   
${\cal O}$. But the rewind time is the same for objects at any  
distance $d$, and is always $\delta t=1/H$. Points further away are   
moving faster, and therefore crossed their greater distance from  
${\cal O}$ in the same time.   
Hence, at a time $\delta t=1/H$ into the past, the whole   
observable Universe was ejected from point ${\cal O}$.   
But ${\cal O}$ can be any point. Therefore the whole  
Universe started from a single point, in a big explosion: the Big  
Bang. Gravity complicates, but does not alter this argument.  
\begin{figure}  
\centerline{\psfig{file=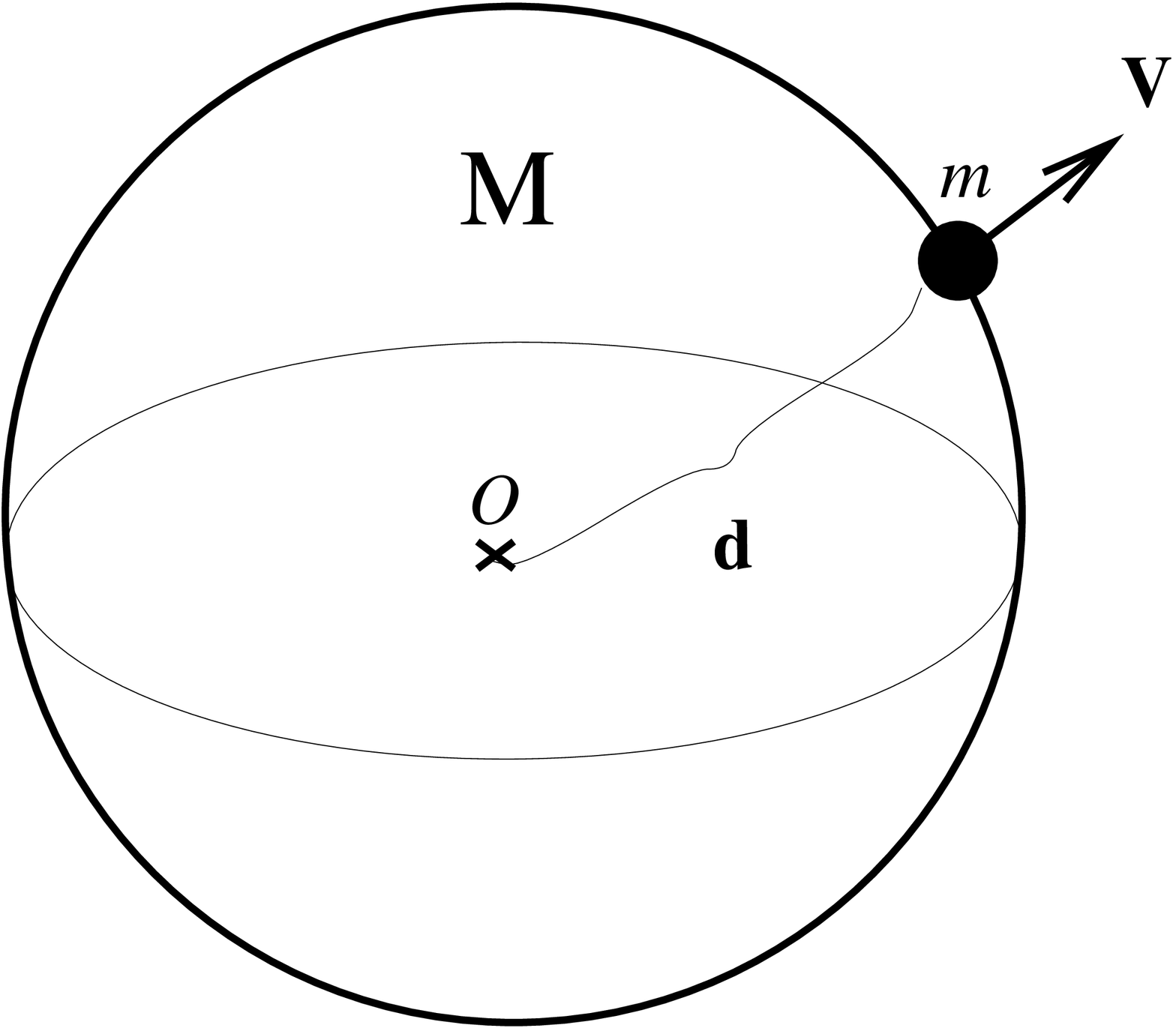,width=6 cm,angle=-90}}  
\caption{}  
\label{fig0a}  
\end{figure}

We shall now include gravity, using O-level algebra. The lazy reader  
may skip this effort: the above already allows pretentious statements about 
creation to be made. Let us again   
consider the perspective of a point ${\cal O}$, and imagine   
a small test particle with mass $m$ at distance ${\bf d}$.  
The gravitational force on the mass $m$ is determined by the mass   
$M$ inside a sphere centred at ${\cal O}$ and with radius $d$ (see   
Fig.~\ref{fig0a}). If $\rho$ is the mass density of the Universe,   
this mass is $M={4\over 3} \pi d^3\rho$. Energy conservation requires that   
\begin{equation}   
-{GMm\over d}+{1\over 2}mv^2=C   
\end{equation}   
(where $v$ is $m$'s velocity) and Newton's law implies that   
\begin{equation}   
m\dot v=-{GMm\over d^2}   
\end{equation}   
where $G$ is the gravitational constant. If we label particle  
$m$ with a comoving coordinate ${\bf l}$, then we may write its  
position as ${\bf d}(t)=a(t){\bf l}$. We call $a$ the expansion  
factor of the Universe. The particle velocity is then ${\bf v}  
={\dot a\over a}{\bf d}$ and its acceleration is ${\dot{\bf v}}=  
{\ddot a\over a}{\bf d}$. With these rearrangements we can derive  
the Friedmann equations:  
\begin{eqnarray}  
{\left({\dot a\over a}\right)}^2&=&{8\pi G\over 3}\rho -{Kc^2\over a^2}  
\label{fried1}\\  
{\ddot a\over a}&=&-{4\pi G\over 3}{\left(\rho+3{p\over c^2}\right)}  
\label{fried2}  
\end{eqnarray}  
in which $K=-2C/(ml^2 c^2)$. In the second equation we have included   
an extra term in $p$, the pressure in the Universe, which does  
not follow from the Newtonian argument.   
This is a correction imposed by general   
relativity even in the Newtonian limit. In Relativity the gravitational  
mass is not given by the mass density $\rho$ alone. The pressure  
contributes as well. We may regard $\rho + 3p/c^2$ as the  
active gravitational mass density of the Universe. This subtlety  
will be of great importance later on.

We see that there is still a Big Bang, even when gravity is taken  
into account. The Friedmann equations give $a\propto t^{2/3}$ for  
a Universe with no pressure (dust), as $t\rightarrow 0$, regardless  
of the constant $K$. The early Universe is in fact filled with radiation,   
for which $p={1\over 3}\rho c^2$. In this case, $a\propto t^{1/2}$  
as $t\rightarrow 0$, for all $K$. In either case we see that   
$a\rightarrow 0$ as $t\rightarrow 0$, that is we have a Big Bang.  
\begin{figure}  
\centerline{\psfig{file=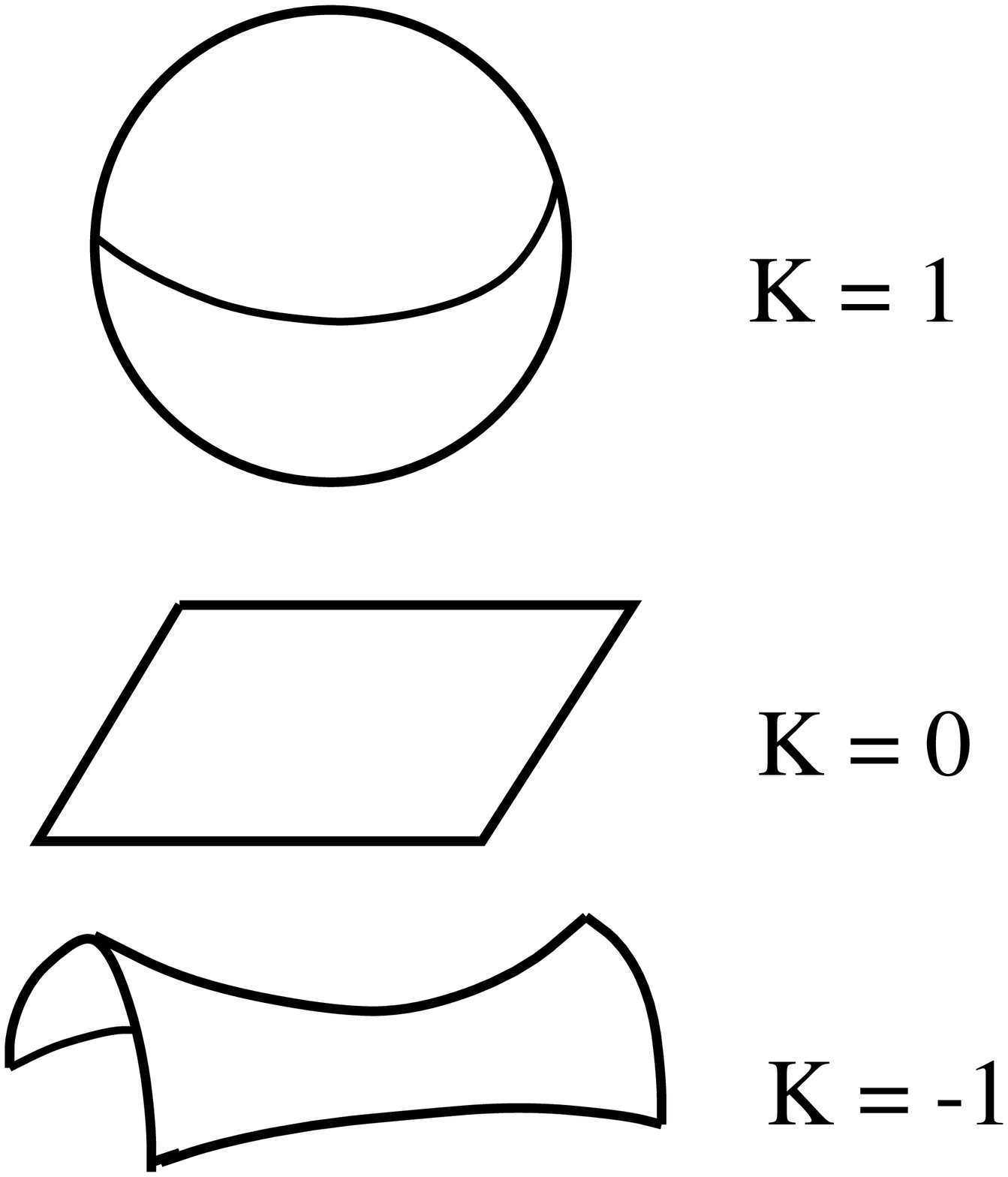,width=6 cm,angle=0}}  
\caption{}  
\label{fig0ab}  
\end{figure}   
  
The previous argument is naturally oversimplified. As $d$ increases   
the recession speed $v$ increases until it eventually approaches the   
speed of light $c$. The Newtonian  argument then breaks down, since special   
relativity invalidates the numerous changes of frame used. Nonetheless  
the Newtonian argument does  
give the right equations. Equations (\ref{fried1}) and (\ref{fried2})  
are Einstein's equations for a uniform and isotropic space-time.  
  
One significant novelty is introduced by relativity. In relativity  
the objects in the Universe are not moving away from each other.  
Rather, they are fixed in space, and space is expanding. Another  
novelty is a better interpretation of the constant $K$ which appears  
in the equations. This constant describes the curvature of the expanding  
space, and $a$ can always be redefined so that  
$K=0,\pm 1$. Given homogeneity, the expanding space can only be  
a 3-dimensional sphere ($K=1$), Euclidean or flat 3-dimensional space  
($K=0$), or a 3-dimensional hyperboloid or saddle ($K=-1$). The 2-dimensional  
analogues are pictured in Fig.~\ref{fig0ab}.

The brightest side of the Big Bang model is the prediction of   
Universal expansion. What set the Universe in motion? The question  
does not make sense. It's like asking what keeps a free particle  
moving, as Aristotelian physicists would do. The cosmological expansion  
is a generic feature of any space-time satisfying Einstein's equations,  
as they were written above. Only a restless Universe is consistent with  
relativity; and that is just what was discovered by observation.  
   
\section{The spooky side of Big Bang cosmology} \label{bbp} 
The Big Bang model is a success. It offers the most minimalistic   
explanation for all the observations currently available. It explains   
the cosmic microwave background. It explains the abundances of the 
lighter elements, through a process called primordial nucleosynthesis.   
It provides an explanation for how structures, such as galaxies,   
form in a Universe which is very homogeneous at early times, indeed at any   
time at very large scales. This is only   
to mention a few striking successes of the Big Bang model. Competitors   
to the Big Bang model, such as the steady state model, lost  
their elegance and predictive power as more and more data accumulated.    
   
In the late 70's, however, it became apparent that  
not all was a bed of roses   
with the Big Bang model. Even though the model proved unbeatable    
when confronting observation, it required a large amount of coincidence   
and fine tuning, which one would rather do without. These difficulties 
are referred to as the horizon, flatness, and Lambda problems, 
which we now describe (see Linde 1990 for a review).

\subsection{Horizons in the Universe}   
Creation entails limitation. Universes marked by a creation event,   
such as the Big Bang Universe, suffer from a  
disquieting phenomenon known as   
the horizon effect: observers can only   
see a finite portion of the Universe. The horizon effect can be   
qualitatively understood from the fact that, since light takes time to    
travel, distant objects are always seen as they were in the past.   
Given that creation imposes a boundary in the past, this means that   
for any observer there must also be a boundary in space.   
A distance must exist     
beyond which nothing can be seen, as one would be seeing objects before    
the creation. Such a boundary is called the horizon.   
  
\begin{figure}  
\centerline{\psfig{file=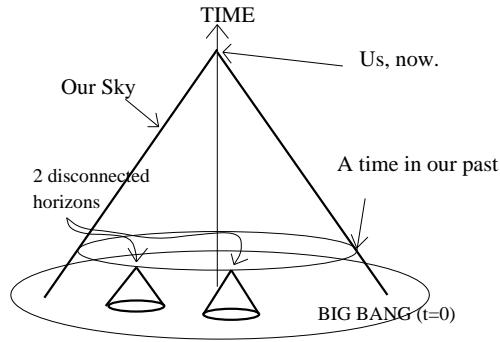,width=6 cm,angle=-90}}  
\caption{Conformal diagram (light at $45^\circ$) showing the  
horizon structure in the Big Bang model. Our past light cone contains   
regions outside each others' horizon.}  
\label{fig1}  
\end{figure}  
 
The existence of horizons is not by itself a problem. The problem 
is that the horizon is tiny at early times. If we ignore 
expansion effects, the current horizon radius is 15 billion light years, 
corresponding to our age of 15 billion years. When the Universe was 
100,000 years old the horizon radius was only 100,000 light years.  
If we look far enough we can see the 100,000 year old Universe. 
As Fig.~\ref{fig1} shows, we should be able to see many regions 
which were outside each other's horizons at that time. 
 
The celebrated cosmic microwave background radiation is, in fact, a 
glow emitted by the Universe when it was 100,000 years old. One can 
show that a horizon region at this time subtends in the sky  
an angle of about a degree, the size of the moon.  
We can see many of these regions, and they all agree to have the  
same temperature to a high degree of accuracy.  
 
The horizon problem is  our ability to see disconnected horizons 
in our past, and the fact that these horizons are seen to share    
the same properties, such as temperature or density. The horizon effect    
prevents any causal equilibration mechanism from explaining this remarkable   
coincidence. In a sense the horizon problem  is really 
a homogeneity problem:  the uncanny homogeneity of the Universe 
across causally disconnected regions.

Expansion complicates this reasoning a bit, but not enough to 
solve the problem in a Big Bang Universe. When we discuss the inflationary 
Universe we shall include the effects of expansion into the discussion, 
and show how they may be used to solve this riddle.

\subsection{Walking on a tightrope without falling off}   
The first Friedmann equation, (\ref{fried1}), shows that there are two  
contributions to expansion. On its right hand side we see that matter  
and, if present, curvature contribute to expansion.  How does their 
relative importance evolve in time? 
  
The two Friedmann equations may be combined into a single equation  
\begin{equation}\label{cons1}  
\dot\rho+3{\dot a\over a}{\left(\rho+{p\over c^2}\right)}=  
0  
\end{equation}  
which represents energy conservation. This implies that pressureless matter  
($p=0$) is diluted as $\rho\propto 1/a^3$, the dilution rate  
corresponding to volume expansion. Radiation ($p={1\over 3}\rho c^2$)  
is diluted as $\rho\propto1/a^4$, the extra factor corresponding   
to the fact that the radiation pressure is doing work as the Universe  
expands. Eqn. (\ref{fried1}) then 
shows that the ratio between the contribution to expansion  
due to curvature, and that due to matter (radiation), increases with  
$a$ ($a^2$). This means that the flat model is unstable!  
This conflicts with the fact that we are now close to flatness.  
Observations show that the  
contributions to expansion due to matter and curvature are at most  
of the same order of magnitude. How have we managed not to fall off  
the tightrope of flatness? This is the flatness problem.  
  
To put numbers into the problem, we know that the Big Bang Universe   
has been expanding since the so-called Planck time, $t_P=10^{-42}$ sec, 
when gravity   
became classical. We can work out that since then the expansion factor   
has increased by $10^{32}$, or thereabouts, leading to a growth in any  
deviation from flatness by around $10^{64}$ since then. This means fine  
tuning the curvature contribution at Planck time by 64 orders of magnitude.  
  
Falling off the flatness ridge is disastrous. Staring at the  
first Friedmann Equation (\ref{fried1})  
we see that closed models ($K=1$) expand more slowly 
than flat models, but open models ($K=-1$) expand  
faster. We can follow the evolution  
of non-flat Universes into their curvature dominated epochs.  
Expansion in closed models keeps slowing down relative to flat models,   
until eventually expansion comes to an halt. The curvature term  
cannot be bigger than the matter term (since 
the left hand side of (\ref{fried1})  
must be positive). Therefore recollapse starts, retracing expansion's 
steps, until the Universe ends in a Big Crunch.   
Open models expand ever more rapidly compared to flat models. Therefore the  
curvature term keeps increasing until the matter is irrelevant.   
But this means that the Universe is destined to become totally empty.  
If the contribution to expansion from curvature is not negligible 
at Planck time, one or the other of these two tragedies  
would have occurred within a few Planck times.  
  
Only flat models offer a reasonable  
model for the Universe as we see it. But as we saw, they are unstable, 
the tiniest trace of curvature sufficing to derail them.

\subsection{Einstein's greatest blunder}   
Self-flagellation has played an important role in modern science.  
At the start of the 20th century there was no  
evidence for cosmological expansion. Relativity predicts expansion,  
with one exception: a closed Universe dominated by a cosmological  
constant. The cosmological constant represents the energy of the vacuum,  
and it was introduced by Einstein to ward off expansion. When  
Hubble discovered expansion, a few years later, Einstein  
bitterly regretted having introduced the cosmological constant,  
thus missing yet another theoretical prediction for a major experimental  
discovery. He called the cosmological constant  
``the biggest blunder of my life''.

The cosmological constant may be seen as an extra term one adds  
to curvature in Einstein's equations. This term is usually represented  
by $\Lambda$. It may be reinterpreted as an extra fluid pervading the  
whole Universe, with pressure   
$p_\Lambda=-\rho_\Lambda c^2$, and mass density $\rho_\Lambda=\Lambda c^2/  
8\pi G$. The cosmological constant is the stuff the vacuum is made of.  
  
The cosmological constant has a very negative pressure,  
that is, it is very tense stuff. Inserting this fact into Eqn. (\ref{cons1})  
leads to $\rho_\Lambda={\rm const}$. The vacuum does not get  
diluted by expansion! This is because expansion is doing work against  
the $\Lambda$ tension. Therefore at the same time expansion dilutes  
the energy density in $\Lambda$, it transfers energy into it, via this work.  
  
We see that any traces of the cosmological constant would immediately   
dominate the Universe. Defining the ratio between the energy density  
in normal matter and in $\Lambda$ reveals that   
$\epsilon_\Lambda=\rho_\Lambda/\rho$ grows like $a^3$ for matter,  
and like $a^4$ for radiation. This means a growth by 128 orders of  
magnitude since Planck time, when we know the Universe expansion  
must have started.   
  
We have another, even thinner,  tightrope to walk on.

\section{God on amphetamine}  \label{infl} 
In the end, history flushed Einstein's greatest blunder into  
one of the main paradigms of modern cosmology: inflation  
(Guth 81, Linde 82, Albrecht \& Steinhardt 82, Linde 83).  
Inflation is a period in the early Universe during which the dominant  
energy contribution is the vacuum energy. Inflation is a brief  
affair with the cosmological constant.   
  
Inflation is a way of switching on the cosmological constant, and then  
letting it decay into ordinary matter. The trick is played by a field,  
called the inflaton field. When the inflaton is switched on it dominates  
all other forms of matter, in the catastrophe described above. However  
this catastrophe brings luck. Integrating the Friedmann equations with  
$p=-\rho c^2$ leads to $a\propto e^{Ht}$, where the Hubble constant $H$  
is now indeed a constant. We have exponential expansion. The Universe 
therefore inflates, and this,  
as we shall see, is enough to solve the flatness and horizon  
problems.  
  
Inflation may be achieved with stuff less extreme than a temporary  
cosmological constant. In fact $\rho+3p/c^2<0$ is the generic condition   
for inflation. It means  
that the gravitational mass of the Universe is negative. For  this reason  
the cosmological expansion accelerates instead of decelerating ($M<0$  
in Fig.~\ref{fig0a}). More precisely $\ddot a>0$, as we can see  
from the second Friedmann Equation (\ref{fried2}). A period of 
inflationary expansion  
is sometimes also called ``superluminal expansion''.  

\subsection{Opening up horizons}   
In our discussion of the horizon problem we neglected expansion. 
Let us now refine the argument. The horizon size is the distance  
travelled by light since the Big Bang. But is this really one 
light year in a one year old Universe? Travelling in an expanding  
Universe entails a surprise: the distance from the departure point is  
larger than the distance travelled.  
This is because expansion keeps stretching the distance   
already travelled. Imagine a cosmic motorway, realized if the Earth 
were expanding very fast. Then a trip from London to Durham   
might show on the speedo that 300 miles were travelled,  
whereas the actual distance   
between the two places at the end of the trip would be 900 miles.   
Similarly in a 15 billion year old Universe, light would have travelled   
15 billion light years since the Big Bang. However the distance to its   
starting point would be roughly 45 billion light years, the  
current size of the  horizon.   
 
This subtlety does not change the essence of the discussion of  
the horizon effect in Big Bang models.  But inflation builds upon this 
subtlety. With super-luminal expansion the distance travelled by light 
since the start of inflation becomes essentially infinite. Under  
amphetaminic expansion  
light travels a finite distance, but expansion works ``faster than light'',  
infinitely stretching the distance from departure.

Therefore inflation opens up the horizons. The whole observable   
Universe nowadays was, before inflation, a tiny bit of the Universe  
well in causal contact. This was then blown up by a period of inflation.   
We have solved the horizon problem.

\subsection{The valley of flatness}   
If we insert a cosmological constant into the flatness problem  
argument, we find a pleasant reversal of the situation.   
Now the contribution to expansion due to matter (which is $\rho_\Lambda$)   
remains constant, whereas the contribution due to curvature decays   
like $1/a^2$. The ratio between curvature and matter contributions  
now decreases like $1/a^2$ instead of increasing like $a^2$.  
Flatness becomes a valley, rather than a ridge.   
  
Because the expansion  
factor is increasing exponentially, within a very short time  
any deviation from flatness becomes infinitesimally small.  
At the end of inflation the contribution to expansion from curvature  
is smaller than $10^{-64}$. We have achieved the fine  
tuning required to survive the Big Bang flatness tightrope. Inflation  
provides the primordial balancing pole to allow us to walk the tightrope  
without falling off.  
  
\subsection{The end of the nothing}  
At the end of inflation   
the inflaton field decays into radiation, in a process  
known as reheating. The normal course of the Big Bang resumes.  
But the worst Big Bang nightmares have been staved off.   
It is no longer a coincidence that the Universe is homogeneous across so many  
disconnected horizons. All these separate horizons went to the  
same nursery school. The instabilities of the sensible brand of Big Bang  
models (the flat ones) are no longer a concern. A period  
of inflation finely tuned the Universe. It gave it the stability  
at birth required for the Universe to cope with its  
``instabilities'' in later life.

The only problem inflation does not solve is of course the $\Lambda$  
problem. Inflation is built upon it. If in addition to the   
inflaton effective $\Lambda$, which turns on and off, there is a genuine  
cosmological constant, this will still be present at the end of inflation.  
The energy densities for the two $\Lambda$s remain constant, 
and therefore at fixed ratio, during inflation. 
Hence a genuine $\Lambda$ would still threaten to dominate the   
Universe at any time after inflation. Inflation  
does not provide the fine tuning required to solve the $\Lambda$  
instability of the Big Bang.

\section{Is there life before the Big Bang?} \label{pre-bb}  
There have been several attempts to solve the Big Bang riddles  
by plunging into the Planck time, $t_P=10^{-42}$ sec, before which  
the temperatures in the Universe are so high that gravity, and   
therefore the evolution of the Universe, becomes dominated by quantum  
effects. Perhaps the most challenging approach is quantum cosmology,  
an attempt to describe the Universe with a wave function, hopefully   
subject to a Schroedinger-type Equation. We shall not describe this  
side of the story. Instead we will  present a few ideas suggesting  
that before the Big Bang there may have been another classical period in the   
life of the Universe. In this previous incarnation one seeks solutions  
to the cosmological puzzles.   
  
\begin{figure}  
\centerline{\psfig{file=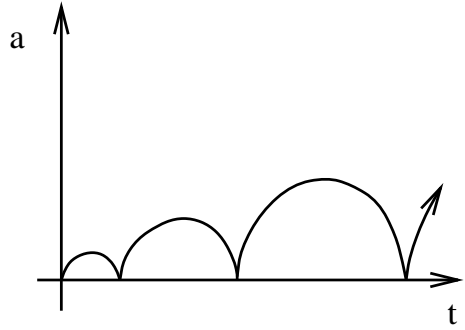,width=6 cm,angle=-90}}  
\caption{The scale factor evolution in the bouncing Universe.}  
\label{bounce}  
\end{figure}

Historically the first such attempt was the Bouncing  
Universe (see Zeldovich and Novikov 83).  
Closed Universes expand to a maximum size, and then recollapse, eventually  
reaching a Big Crunch. What if the Crunch bounced into a Bang?  
This cannot be achieved classically, but may be possible due to 
quantum effects, although this remains a speculation.  
In Fig.~\ref{bounce} we plot the typical evolution of the scale factor   
$a$ in such models. The maximum size of the Universe is related to 
its entropy. The second law of thermodynamics then requires that the  
bouncing Universe gets bigger in each cycle.   
  
A bouncing Universe does not have an horizon. To see this let us  
ask the question: can a light  
ray in a closed Universe ever get back to its starting point?  
Are there Magellanic photons in a spherical Universe? 
The answer is yes: if a ray sets off at the Big Bang,  
it travels around the Universe  
and gets back to the departure point at the Big Crunch. Hence after the first 
cycle all points  have been in causal contact. It is only if we are 
unaware of the cycles preceding our own that we may infer  
a horizon problem. 
 
However it turns out that even though we have solved the horizon  
problem, we have not solved the homogeneity problem. Ensuring causal  
contact between the whole observable Universe allows for an  
equilibration mechanism to homogenise the whole Universe, but  
such a mechanism must still be proposed and be efficient enough.  
No such mechanism seems to be present in Bouncing Universes.  
  
\begin{figure}  
\centerline{\psfig{file=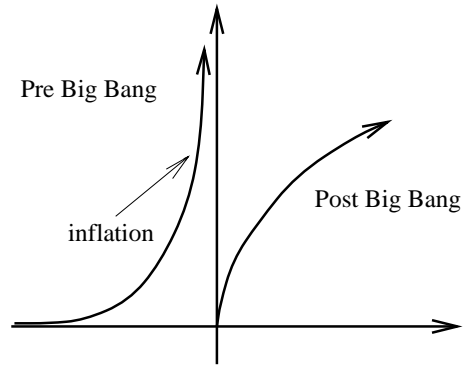,width=6 cm,angle=-90}}  
\caption{The scale factor evolution in Pre Big Bang cosmology}  
\label{prebb}  
\end{figure}

A more modern way to explore life before the Big Bang was recently   
inspired by string theory (Veneziano 97, Gasperini and Veneziano 93).  
In string theory there are a number of duality  
symmetries, typically involving transforming big things 
into small things,  
and strong coupling into weak coupling. In the context of cosmology  
this is reflected in a scale factor transformation of the form  
$a(t)\rightarrow a^{-1}(-t)$. This permits the extension of the history  
of the Universe into times before the Big Bang, times $t<0$.  
For such times the solution dual to the radiation Post Big Bang solution  
is $a\propto (-t)^{-1/2}$. We have accelerated expansion: $\ddot a>0$,  
and therefore we have inflation. The typical time evolution  
of the scale factor is described in Fig.~\ref{prebb}.  
  
The PreBig Bang is really another way of getting inflation.  
But this time inflation occurs before the Big Bang.  
The Pre Big Bang scenario solves the horizon and flatness problems   
in the same way as inflation, but the timing is completely different.

Another aspect of the duality transformation assisting   
Pre Big Bang cosmology involves a field called the dilaton $\phi$.  
The dilaton appears in all attempts to derive low energy limits  
to string theory. It plays the role of a coupling constant for  
all interactions, or rather the couplings are given by, eg,
$G=e^{\phi}$.   
The duality transformation described above requires a transformation  
upon the dilaton of the form $\phi\rightarrow \phi +6\log (a)$. Hence  
the radiation dominated Big Bang solution, with constant dilaton  
$\phi=\phi_0$ (that is with stabilised constants)   
is mapped in the Pre Big Bang epoch  into a solution of the form   
$\phi=\phi_0-3\log (-t/t_0)$. When $t\rightarrow -\infty$ we have  
$\phi\rightarrow -\infty$, and so the generic coupling constant 
$G\rightarrow 0$. This means the Pre Big Bang Universe emerges  
from an epoch in which the interactions were switched off.   
  
The overall picture is that the Universe starts from the  very  
weak coupling regime, evolving into strong coupling.   
Interactions switch on.  In the   
process inflation is also triggered, solving the riddles  
of the Big Bang. Deep in the strong coupling regime  
string theory effects become important, and lead the Universe into the   
Post Big Bang stage. We do not know what happens in the Bang.  
However we hope that the duality transformations  
assisting string theory will be enough to perform the mapping   
between these two stages in the life of the Universe.

\section{Quick-light}  \label{vsl} 
  
The special theory of relativity has dominated 20th century physics.   
More than the general theory, the special theory has become part  
of the fabric of physics. Special relativity has been successfully  
combined with quantum mechanics, to striking effect.  
Quantum field theory emerged from the union, with an array of  
spectacular predictions leading to modern particle physics. Good 
examples are the discovery of new particles and  
antiparticles, the electroweak theory and the prospect of unification  
(and the crucial idea of spontaneous symmetry breaking), as well 
as  all sorts  
of high precision quantitative predictions concerning interactions   
and their cross-sections.   
  
Central to special relativity is the idea that the speed of light $c$ is  
a constant. Regardless of the speed of the emitting or observing  
object, light moves at the same speed: about 300,000 Km sec$^{-1}$.  
Nothing can travel faster than light.  
The invariance of $c$ imposes a symmetry group on physics, called  
Lorentz symmetry. Only if space and time transform in a specific  
manner between different observers can the speed of light be the same  
for all of them. The implications of the Lorentz transformation  
are immensely popular. The (Lorentz) contraction of moving bodies,  
time dilation and the twin ``paradox'' are now well known to   
everyone.   
  
\subsection{VSL}  
But what if the speed of light were to change during the lifetime of the  
Universe? ``Varying constant'' theories have been proposed, starting  
with Dirac's idea of varying the gravitational constant $G$.   
In attempting to explain the constants of nature  
one should allow them to vary, and then see if a physical mechanism  
can be found which  
crystallises their values into fixed quantities. If so we may hope that  
these  values are the ones we observe. This project   
has not been terribly successful. But as a byproduct it has left us  
with great insights into what physics would be like   
if indeed the constants of nature were variables.  
  
A good example is the Jordan-Brans-Dicke theory, in which 
the gravitational constant $G$ is a variable.  
In this theory the gravitational constant is the result of the matter  
content of the Universe.  As the cosmic density changes, $G$ changes 
as well. Such theories have led  
to interesting cosmological models, and attempts to solve Big  
Bang riddles with them have been made, albeit unsuccessfully.  
Another example is the theory proposed by Bekenstein, in which  
the electron charge $e$ is a field. String theories predict that all  
charges are in fact variable, and related to a single field, the dilaton  
field.  
  
Varying speed of light  
(VSL) is based on a similar exercise applied to $c$ 
(Moffat 93, Albrecht \& Magueijo 99, Barrow 99).  
In the simplest implementation of VSL, $c$ 
drops in a sharp phase transition in the Early Universe. Light 
was much faster in the Early Universe.  
 
There is an element of criminal negligence in VSL. In VSL  
all observers at the same point, at the same time, but possibly 
moving relative to each other, see the same $c$.  
Again nothing can travel faster than light. However Lorentz symmetry  
is broken.  
And once this happens we are in the dark. Lorentz symmetry  
has been the guiding principle used to set up all new theories 
in the 20th century.  If we discard it, what new guideline can 
we adopt? 
  
We postulate a principle of  
minimal coupling. This means simply to replace  
$c$ by a field $c(t,{\bf x})$ wherever it occurs in selected  
laws.  
Such a minimal coupling principle guided the construction of other  
``varying constant'' theories. It ensures that nothing new happens  
when the ``varying constants'' are kept fixed. It  
also ensures that minimal  
changes are introduced when ``varying constants'' do vary.  
 
Minimal coupling cannot be consistently applied in all laws.  
We decided to apply it 
to the field equations: in the case of gravity, to Einstein's  
equations. Curvature is not affected by VSL, and the way matter generates  
curvature is the same as before. In some loose sense   
we have General Relativity without  
Special Relativity.  
  
In the context of cosmology this means  
simply replacing $c$ by a variable $c(t)$ in Eqns.~($\ref{fried1}$) and  
(\ref{fried2}). The matter content in the Early Universe is still  
relativistic, so $a\propto t^{1/2}$. We don't have   
superluminal expansion.

\subsection{Quick-light years}  
\begin{figure}  
\centerline{\psfig{file=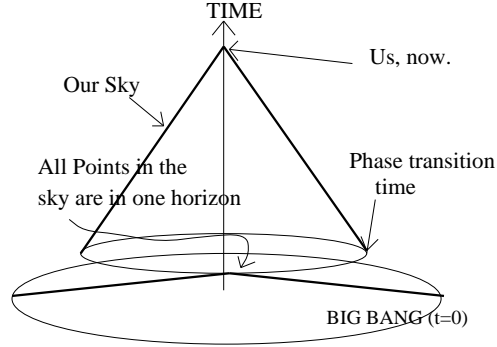,width=6 cm,angle=-90}}  
\caption{Diagram showing the horizon structure in a model  
in which at time $t_c$ the speed of light changed from $c^-$  
to $c^+\ll c^-$. Light travels at $45^\circ$ after $t_c$  
but it travels at a much smaller angle with the space axis before  
$t_c$. Hence it is possible for the horizon at $t_c$ to be much  
larger than the portion of the Universe at $t_c$ intersecting our   
past light cone. All regions in our past have then always been   
in causal contact.}  
\label{fig2}  
\end{figure}  
It is immediately obvious that quick-light in the early Universe solves  
the horizon problem. Look at Fig.~\ref{fig2}, in which we redrew  
Fig.~\ref{fig1} assuming the speed of light was much larger in the  
Early Universe than it is nowadays,  
dropping to its current value in a sharp phase transition at $t=t_c$.  
We don't need to play tricks with expansion in order to   
establish causal contact between the whole observable Universe.  
Even without expansion (so that $d_h=ct$ when $c$ is constant),   
the quick-light pervading   
the Early Universe would have been enough to connect the  
whole observable Universe.  
  
Suppose that the transition happened when the Universe was 1 
year old. The horizon was then one quick-light year across, 
easily bigger than 15 billion normal-light years, if quick-light 
is fast enough. 
If such a phase transition occurred at Planck time ($t_c=t_P$), 
then light would need to have been  $10^{32}$ times faster than  nowadays, 
to solve the horizon  problem.   
  
\subsection{Another valley for flatness}   
Energy conservation appears in relativity  
as a consistency condition for Einstein's equations.  
For instance, before Eqns.~(\ref{fried1}) and  
(\ref{fried2}) can be solved, one must satisfy conservation   
Eqn.~\ref{cons1}, as the latter is implicit in the former two.  
This is only true if $c$ is a constant. If $c$ is allowed to vary  
then combining Eqns.~(\ref{fried1}) and  
(\ref{fried2}) leads to   
\begin{equation}\label{cons2}  
\dot\rho+3{\dot a\over a}{\left(\rho+{p\over c^2}\right)}=  
{3Kc^2\over 4\pi G a^2}{\dot c\over c}  
\end{equation}  
implying lack of energy conservation.

It is not surprising that energy conservation must be violated in such  
a theory. Conservation laws are the result of symmetries: that is the  
modern way to look at it. For instance conservation of angular momentum  
is a consequence of isotropy, the symmetry according to which the Universe  
looks the same in all directions. Energy conservation results from invariance  
under time-translations: the laws of physics are the same at all times.  
This is clearly not the case if the speed of light changes: the laws  
of physics change fundamentally as the speed of light changes. Therefore  
there must indeed be violations of energy conservation if $\dot c/c\neq 0$,  
as shown by Eqn.~\ref{cons2}.

Lack of energy conservation pays its dividends. For a given  
expansion rate, $\dot a/a$, we can diagnose the geometry of  
the Universe by comparing its density with the density corresponding  
to the flat case  $\rho_c={3\over 8\pi}{\dot a\over a}$ (see 
Eqn.~\ref{fried1}). The density $\rho_c$ is called  
the critical density, and if the density is supercritical, $\rho>\rho_c$,  
the Universe is closed; 
if the density is subcritical, $\rho<\rho_c$, the Universe is open.  
Now let us stare at   
Eqn.~\ref{cons2}. We see that if the speed of light decreases  
($\dot c/c<0$), then matter is created if the Universe is open ($K=-1$ 
and $\rho<\rho_c$),  
but disappears if the Universe is closed ($K=1$ and $\rho>\rho_c$).  
There is no creation or anihilation if the Universe is flat ($K=0$ 
and $\rho=\rho_c$). Hence VSL creates matter if we are subcritical,  
subtracts it if we are supercritical. Again we have produced a valley  
for flatness.  
  
It can be shown that a drop in $c$ by 32 orders of magnitude at Planck time  
would provide sufficient fine tuning for a flat Universe to be seen  
nowadays.  
  
\subsection{Exorcising the nothing}  
Finally VSL solves the cosmological constant problem. Einstein introduced 
$\Lambda$ into his equations as an extra geometrical term.  
However the dynamical importance of $\Lambda$ can only be inferred 
when we reinterpret it as a fluid, with a density which remains  
constant under expansion, and with $p_\Lambda=-\rho_\Lambda c^2$.  
The density of this fluid is $\rho_\Lambda=\Lambda c^2/(16 \pi G)$,  
and because $\rho_\Lambda\propto c^2$,  
 we see how a drop in $c$ reduces the dynamical 
significance of $\Lambda$. If $c$ drops by   
more than 64 orders of magnitude at Planck time,  
indeed $\rho_\Lambda\ll \rho$ nowadays. We have exorcised  
vacuum domination.  
  
 
Celebrations of this triumph were interrupted by disturbing claims  
for observational evidence that the cosmic expansion is accelerating, 
$\ddot a>0$ (Perlmutter \& al 1998). This implies that $\Lambda$  
is still with us, and about to dominate the Universe.  
We are about to enter a period of inflation! 
 
This is horrifying. All galaxies will soon recede  
away from us so fast that we will not be able to see them.  
We will soon be confined to 
our galaxy-island, with nothing but the $\Lambda$-vacuum to keep  
us company, in cosmic loneliness.  
We will end up in an island Universe, as Kant envisaged. 
Explaining why $\Lambda$ is only now about to dominate the Universe 
is an outstanding challenge. Why now?  Why not immediately after the Planck 
time? Why not never?  
 
As this review goes to press one of the authors is suffering from  
insomnia due to this humiliating riddle.

\section{An appraisal of current cosmology} \label{conc} 
In this review we provided a rather diluted version of a very  
technical field. We described how the Big Bang model  
converted cosmology into a successful science. We showed how its  
riddles have provided insights into theories of the very Early Universe,  
when the Big Bang must be replaced by something more fundamental.  
We described three classes of models. Inflation is now a paradigm, 
Pre Big Bang models are a popular tentative idea, while 
VSL theories are outright  
speculation. In the words of a distinguished 
Cambridge Professor  ``VSL is a step back from relativity''.  
  
In order to make this review more   
accessible, we highlighted the least technical  
aspects of the Big Bang riddles. 
This necessarily distorts the field. Perhaps the biggest riddle of  
all is the emergence of structure in a Universe known to be very homogeneous  
at Early times. All the above theories can answer this  
riddle, but in ways too technical for a light hearted review like this one.  
  
Nonetheless structure formation is really the testing ground 
where experiment may  
one day decide between all these ideas. We can measure the properties  
of galaxy clustering, and also the power spectrum in the cosmic microwave  
background (CMB) anisotropies. Theories of the Early Universe make 
very different predictions for  
these observations. A new generation of satellite CMB  
experiments, plus new galaxy surveys, leave us at a threshold. In  
the 21st century it could well be decided which if any of the above  
ideas is correct.  
  
The most exciting possibility is of course that all the current ideas  
are proved wrong. For that reason we believe that this is a bad time to 
adopt a dogmatic view in cosmology. Instead,  
we should try out as many new ideas as  
possible. Who knows, even so they may still all be wrong.   
We advocate promiscuity in science.

\vspace{0.5cm} 
{\bf Acknowledgements} JM is indebted to 
Andy Albrecht, John Barrow, and John Moffat, for shaping his 
views on the cosmological problems. We thank the referee 
and David Sington for forcing us to clarify the more obscure 
parts of the text. We finally acknowledge financial support 
from  the Royal Society (JM) and PPARC (KB).

\end{document}